\documentclass[10pt,twocolumn,letterpaper]{article}

\usepackage{cvpr}
\usepackage{times}
\usepackage{epsfig}
\usepackage{graphicx}
\usepackage{amsmath}
\usepackage{amssymb}
\usepackage{pifont}
\newcommand{\xmark}{\ding{55}}%
\newcommand{\newpara}{\vspace{4pt}}%

\usepackage[breaklinks=true,bookmarks=false]{hyperref}

\cvprfinalcopy 

\def\cvprPaperID{****} 

\begin{document}

\title{Naver at ActivityNet Challenge 2019 - Task B Active Speaker Detection (AVA)}

\author{Joon Son Chung\\
Naver Corporation\\
South Korea\\
{\tt\small joonson.chung@navercorp.com}
}

\maketitle

\begin{abstract}
This report describes our submission to the ActivityNet Challenge at CVPR 2019. We use a 3D convolutional neural network (CNN) based front-end and an ensemble of temporal convolution and LSTM classifiers to predict whether a visible person is speaking or not. Our results show significant improvements over the baseline on the AVA-ActiveSpeaker dataset.
\end{abstract}

\section{Introduction}

Multimodal speech perception has received increasing attention in recent years, with significant breakthroughs in audio-visual methods using deep learning~\cite{Afouras19,Afouras18,assael2016lipnet,ephrat2018looking,noda2015audio}. For many applications, it is important to be able to determine which of the visible people are speaking at any given time. To this end, the new AVA-ActiveSpeaker dataset~\cite{roth2019ava} has been an important contribution to the field since it enables deep-learning based active speaker detection (ASD) models to be trained with full supervision. This report gives a brief overview of the dataset, and describes the method that underlies our submission to the challenge.

\subsection{Datasets}

The method is trained on the AVA-ActiveSpeaker dataset. The dataset consists of training, validation and test sets, and the splits are given in Table~\ref{tab:data}. Ground truth labels are given for the training and validation sets.

\begin{table}[ht] 
\setlength{\tabcolsep}{10pt}
\begin{center}
\begin{tabular}{ l r r } 
 &  \\  \hline
 \bf{Set}  & \bf{\# videos} &  \bf{\# frames} \\  \hline
 Train  & 120 &  2,676K \\
 Val  & 33 &  768K \\ 
   Test  & 109 &  2,054K \\  \hline
 
\end{tabular}             
\end{center}
\caption{Statistics of the AVA-ActiveSpeaker dataset}
\label{tab:data}
\end{table}

The dataset is challenging for a few reasons. The durations of speaking segments are extremely short, the average Speaking \& Audible segment being 1.11 seconds. As a result, the system must be able to make accurate detection using only a small number of frames. Existing methods that require the output to be smoothed over several-second time window~\cite{chakravarty2016cross,chung2016out} cannot be effective for this condition.

The dataset also contains a significant number of older videos in which the audio and the video appear to have been recorded separately or significantly out-of-sync. This means that the temporal correspondence between audio and video speech representations~\cite{chung2016out} is not a reliable indicator of whether a person is speaking.

\section{Model}

The active speaker detection model consists of front-end feature extractors and a back-end classifier, each of which will be described in the following sections.

\subsection{Front-end architecture}

\begin{figure}[ht]
\begin{minipage}[b]{0.49\linewidth}
  \centering
  \centerline{\includegraphics[width=1\linewidth]{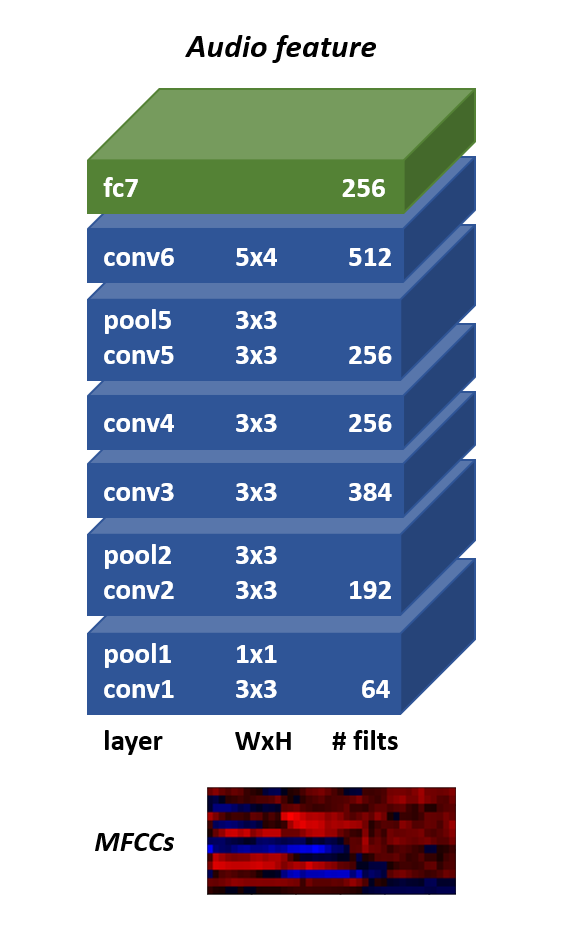}}
  \centerline{(a) Audio stream}\medskip
\end{minipage}
\begin{minipage}[b]{0.49\linewidth}
  \centering
  \centerline{\includegraphics[width=1\linewidth]{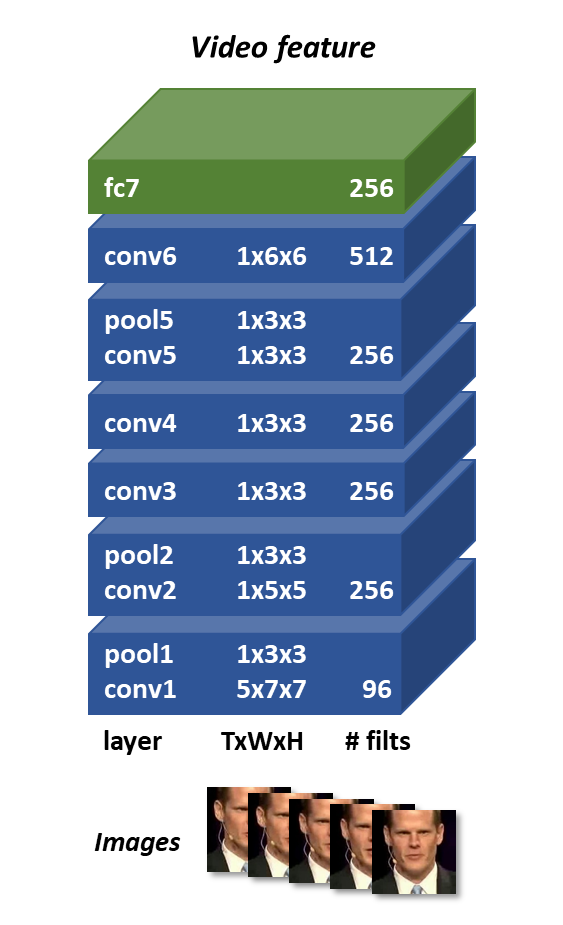}}
  \centerline{(b) Visual stream}\medskip
\end{minipage}
\caption{Front-end architecture for audio and visual encoders}
\label{fig:streams}
\end{figure}

We use pre-trained networks to extract audio and video representations. The encoder networks have been trained on the audio-visual correspondence task using self-supervision on unlabelled videos~\cite{chung2018perfect}\footnote{Model and code available from the author's website.}. 

The video encoder is a convolutional neural network (CNN) that ingests 5 RGB image frames  and outputs a 512-D representation. The architecture is based on the compact but efficient VGG-M network~\cite{Chatfield14}, but with a 3D convolution in the first layer in place of the standard 2D convolution.

The input to the audio encoder is 20 frames in the time direction and 13 cepstral coefficients in the other direction. The encoder  generates a 512-D representation in the same embedding space as the video representation. The detailed architecture is given in Figure~\ref{fig:streams}.

\subsection{Back-end architecture}

\begin{figure}[ht]
\centering
\includegraphics[width=0.9\linewidth]{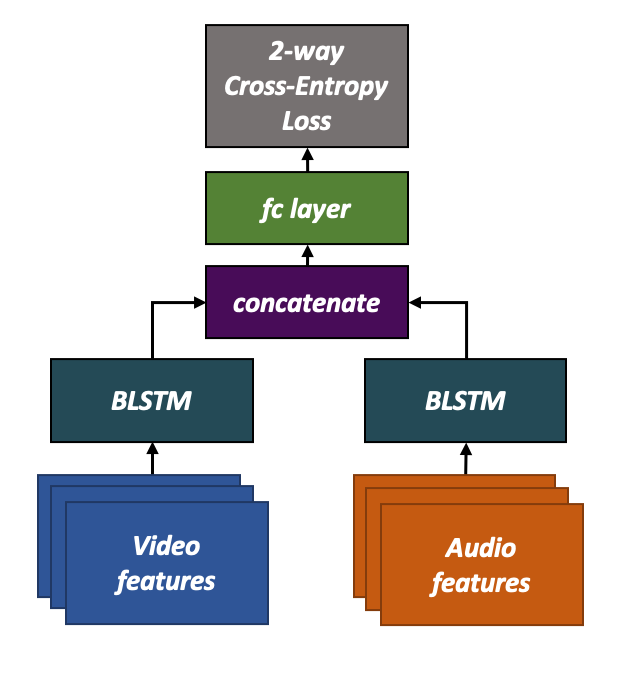}
\caption{LSTM-based back-end classifier. The architecture of the TC-based model is identical, except for the TC layers that replaces the BLSTMs.}
\label{fig:streams}
\end{figure}

Both audio and video encoders ingest 5 video frame (0.2-second) input, moving 1 video frame (0.04-second) at a time. Hence for a $T$ frame input, the output dimensions are $512 \times (T-4)$. Here, we compare two simple back-end classifiers. In our experiments,  $T=9$ is used, but we observed no significant effect on performance for changes to the value of $T$ in the range  $T \in [7,15]$.

\newpara\noindent\textbf{LSTM classifier.} 
The audio and video representations are fed into two separate bi-directional LSTM networks, each of 2 layers and hidden size of 128. The outputs from both networks are concatenated, then passed through a linear classification layer that predicts whether the person is speaking or not. The classifier is trained with the softmax cross-entropy loss.

\newpara\noindent\textbf{TC classifier.} 
Instead of the LSTM layers, the encoder outputs are passed to two temporal convolution layers each with 128 filters. The outputs are also concatenated and passed to the classifier, as in the case of the LSTM classifier.

\newpara\noindent\textbf{Ensemble.} 
In many applications of machine learning, it has been shown that ensemble methods can often perform better than any single classifier~\cite{dietterich2000ensemble}. 
Here, the predictions from both LSTM and TC classifiers are averaged with equal weighting in order to generate the final prediction.

\newpara\noindent\textbf{Smoothing.} 
In order to remove noise in the predictions, the output of the classifiers are smoothed in the temporal domain using median or Wiener filter, both over 0.5-second windows.

\section{Experiments}

\newpara\noindent\textbf{Implementation details.} 
Our implementation is based on the PyTorch library~\cite{paszke2017automatic}
and trained on a single Tesla M40 card with 24GB memory. 
The network is trained 
using the ADAM optimiser~\cite{kingma2014adam} with the default
parameters and a fixed learning rate of $10^{-2}$. To eliminate the effect of bias in the training data, the number of samples for positive and negative classes are balanced in each mini-batch during training.

\newpara\noindent\textbf{Evaluation metrics.} The metric used in this task is the mean Average Precision (mAP). The evaluation code is provided by the organisers of the challenge.

\newpara\noindent\textbf{Results.} The results on the validation set using the different back-end classifiers are given in Table~\ref{tab:res}. 

Our best performing model also achieved mAP of 0.878 on the held-out test set for the challenge. In comparison, the GRU-based baseline model~\cite{roth2019ava} produces mAP of 0.821. 

Qualitative results of the proposed method far exceed existing correspondence-based method~\cite{chung2016out} on this dataset, since it does not rely on precise audio-to-video synchronisation.

\begin{table}[ht] 
\setlength{\tabcolsep}{10pt}
\begin{center}
\begin{tabular}{ l l r } 
 &  \\  \hline
 \bf{Back-end}  & \bf{Smoothing} &  \bf{mAP} \\  \hline
LSTM  & \xmark &  0.851 \\
TC & \xmark &  0.855 \\ 
Ensemble & \xmark &  0.861 \\ 
Ensemble & Median &  0.874 \\  
Ensemble & Wiener &  0.878 \\  \hline
 
\end{tabular}                                
\end{center}
\caption{Results on the AVA-ActiveSpeaker validation set.}
\label{tab:res}
\end{table}

\clearpage
{\small
\bibliographystyle{ieee}
\bibliography{shortstrings,mybib}
}

\end{document}